\documentclass[sigconf]{acmart}

\AtBeginDocument{%
  }

\setcopyright{none}
\copyrightyear{}
\acmYear{2018}
\acmDOI{}

\acmConference[RecTemp Workshop, RecSys’24]{}{Oct 14--18, 2024}{Bari, Italy}
\acmBooktitle{{RecTemp Workshop at RecSys’24, Oct 14--18, 2024, Bari, Italy}}
\acmISBN{}

\usepackage{algorithm}
\usepackage{algpseudocode}
\usepackage{enumitem}
\usepackage{amsmath}
\usepackage{amsfonts}
\usepackage{balance}
\usepackage{caption}
\usepackage{bm}
\setcopyright{none}
\begin{document}

\title{Guided Diffusion-based Counterfactual Augmentation for Robust Session-based Recommendation}

\author{Muskan Gupta}
\email{muskan.gupta4@tcs.com}
\affiliation{%
\institution{TCS Research}
\city{New Delhi}
\country{India}}
\author{Priyanka Gupta}
\email{priyanka.g35@tcs.com}
\affiliation{%
\institution{TCS Research}\city{New Delhi}\country{India}}
\author{Lovekesh Vig}
\email{lovekesh.vig@tcs.com}
\affiliation{%
\institution{TCS Research}\city{New Delhi}\country{India}}

\renewcommand{\shortauthors}{Muskan Gupta, Priyanka Gupta \& Lovekesh Vig}

\begin{abstract}
Session-based recommendation (SR) models aim to recommend top-K items to a user, based on the user's behaviour during the current session. Several SR models are proposed in the literature, however, concerns have been raised about their susceptibility to inherent biases in the training data (observed data) such as popularity bias. SR models when trained on the biased training data may encounter performance challenges on out-of-distribution data in real-world scenarios. One way to mitigate popularity bias is counterfactual data augmentation. Compared to prior works that rely on generating data using SR models, we focus on utilizing the capabilities of state-of-the art diffusion models for generating counterfactual data. We propose a guided diffusion-based counterfactual augmentation framework for SR. Through a combination of offline and online experiments on a real-world and simulated dataset, respectively, we show that our approach performs significantly better than the baseline SR models and other state-of-the art augmentation frameworks. More importantly, our framework shows significant improvement on less popular target items, by achieving up to $20\%$ gain in Recall and $13\%$ gain in CTR on real-world and simulated datasets, respectively. 
\end{abstract}



\maketitle

\section{Introduction}
In today's era, recommendation systems, have become indispensable to numerous applications such as advertising \cite{zhao2021dear}, e-commerce \cite{karimova2016survey}, and streaming services \cite{hasan2018excessive}, transforming how a user interacts with digital platforms. Conventional recommendation systems often generate predictions solely based on the most recent click \cite{hidasi2012fast, xue2017deep}. However, session-based recommendation (SR) models have the ability to enhance user experience, by guiding users toward relevant items based on past interactions. SR models have gained significant attention in recent years. Advance techniques such as Recurrent Neural Networks (RNNs) \cite{hidasi2015session}, attention mechanisms \cite{li2020time, kang2018self,liu2018stamp}, and Graph Neural Networks (GNNs) \cite{wu2018session,yu2020tagnn,he2020lightgcn,fan2019graph} have significantly impacted the performance of the SR models. However, SR models undergo several challenges when deployed in real-world scenarios. 
Given a slate of items/products, users positively interact (click/buy) with a small subset of items leading to sparse user-item interaction data. As a result, an SR model trained on sparse/biased data recommends more popular items to the user, leading to popularity bias. Researchers have attempted to tackle these challenges via normalized representations for item and session graphs \cite{gupta2019niser}, causal inference \cite{gupta2021causer} and data augmentation \cite{gupta2024scm4sr, liu2023diffusion}.
A recently proposed causal inference (CI) based data augmentation approach  generates counterfactual sessions in addition to the observed sessions \cite{wang2021counterfactual,gupta2024scm4sr}. However, these methods rely on generating the counterfactual slate by utilizing the underlying SR model. 
Recent advancements in modern generative models have significantly impacted conceptualization and implementation of SR models \cite{deldjoo2024review}. Researchers \cite{wang2023diffusion,liu2023diffusion,yang2024generate} have explored diffusion models for sequential recommendations  leveraging their impressive generative capabilities. For instance, recent work \cite{yang2024generate} generates the unobserved oracle (unobserved/imaginary) items by modelling the underlying session data distribution using a diffusion model.  Nevertheless, these models struggle to adapt to dynamic user behavior. \par
In this paper, instead of leveraging  diffusion models as recommendation agents, we utilizes the generation capabilities of Guided Diffusion Model to generate a counterfactual/alternate slate of items and leverage a temporal Structural causal model \cite{gupta2024scm4sr} to obtain the counterfactual user response (i.e click/buy) on the counterfactual slate, thus obtaining quality counterfactual sessions.
To this end, we present a guided \textbf{D}iffusion-based \textbf{C}ounterfactual \textbf{A}ugmentation framework for Robust \textbf{SR} models (\textbf{DCASR}) aimed at generating high quality counterfactual sessions. 
We further re-optimize the SR model using both generated counterfactual sessions and observed sessions. We show \textbf{DCASR} outperforms competing models by generating relevant counterfactual sessions that alleviate the popularity bias. \\
The main contributions of this paper are summarized as follows:
\begin{enumerate}[topsep=1pt]
    \item[\textbullet] We propose a novel guided diffusion-based counterfactual data augmentation framework DCASR for session based recommendations. Our framework generates an alternate slate of items by utilizing the generative capabilities of diffusion models.
    \item[\textbullet] We leverage generative diffusion models and causal inference to generate counterfactual samples which are used to further optimize the SR model.
   \item[\textbullet] We conduct experiments on simulated and real-world datasets, demonstrating that our model outperforms current state-of-the-art methods. Additionally, we show that our framework effectively mitigates popularity bias.
\end{enumerate}


\section{Related Work}
In recent years, the use of deep learning based techniques have significantly improved  the  
performance of SR Models \cite{yu2020tagnn, wu2018session,hidasi2015session,liu2018stamp}. Nevertheless, challenges persist when dealing with popularity bias or data sparsity. To address these challenges, researchers have proposed various techniques. NISER \cite{gupta2019niser} tackles these challenges by learning the normalized representations for item and session graphs. CauSeR \cite{gupta2021causer} proposed a generalizable way to handle popularity bias at the data and model level using causal inference. 

Further, various data-augmentation based methods have been proposed to handle popularity bias for SR. \cite{wang2021counterfactual} proposed a data augmentation approach, where less noisy counterfactual samples are generated using adversarial learning. \cite{zhang2021causerec} model the  counterfactual data distribution via contrastive user representation learning. 
\cite{gupta2024scm4sr} proposed a structural causal model based framework to generate counterfactual sessions which are used to update the SR model. Recently, diffusion models have been proposed for data-augmentation for SR.
\cite{wu2023diff4rec} propose a diffusion
model that is pre-trained on recommendation data via corrupting and reconstructing the user-item interactions in the latent space, and the generated predictions are leveraged to produce diversified augmentations for the sparse user-item interactions.
\cite{liu2023diffusion} propose a
diffusion-based pseudo sequence generation framework to fill the gap between image and sequence generation.  Then, a sequential U-Net is designed to adapt the diffusion noise prediction model to the sequence generation task.
\section{DCASR: The proposed Framework}
\begin{figure}[t]
\center
{\includegraphics[scale=0.27]{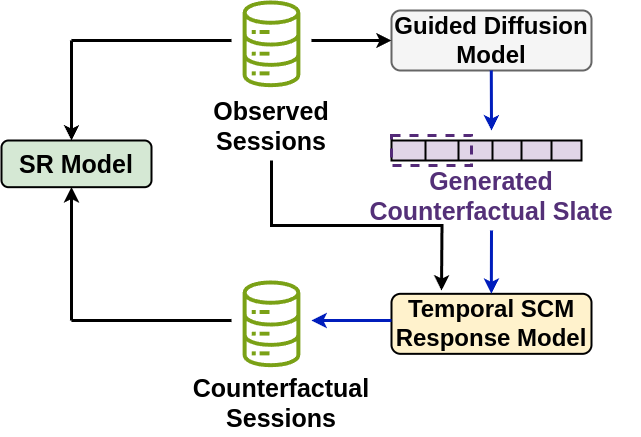}} 
\caption{An overview of our proposed framework. Here black and blue color arrows represent the training and generation phase respectively.}\label{fig:overview}
\end{figure}
In the following section, we introduce the proposed \textbf{DCASR} framework.
The overall structure of our proposed guided diffusion-based counterfactual augmentation framework for SR model is shown in figure \ref{fig:overview}. Our framework comprises of three main blocks: i) SR models, ii) Guided Diffusion model,  and iii) Temporal SCM Response model. In the following sub sections we discuss the details of each block of the framework separately. We first discuss the basic notations corresponding to SR model ($\mathcal{M}_r$) in section \ref{sec:srmodel} . Next we discuss the Guided Diffusion Model ($\mathcal{M}_d$) and Temporal SCM Response Model ($\mathcal{M}_s$) in the subsequent sections respectively. As shown in figure \ref{fig:overview}, we train the $\mathcal{M}_d$ and $\mathcal{M}_s$ models on the observed sessions data. Next, with the aim of generating high quality alternate/counterfactual sessions for a user, we first generate the counterfactual slate of items using $\mathcal{M}_d$ and then generate the response (i.e click/skip) on a counterfactual slate by utilizing the capabilities of a pre-trained $\mathcal{M}_s$ model. Finally, we train the $\mathcal{M}_r$ model, on both the observed and generated counterfactual sessions. The trained $\mathcal{M}_r$ model is finally used for recommendations. We summarize our approach in Algorithm \ref{algo:scm}.
\subsection{SR Model}
\label{sec:srmodel}
Suppose that $\mathcal{S}$ denotes the set of all sessions in the logged data containing user-item interactions (e.g. click/view/order), and $\mathcal{I}$ denotes the set of $m$ items observed in $\mathcal{S}$.
Any session $s \in \mathcal{S}$ is a sequence of item-click events: 
$s = (i_{s,1},i_{s,2},\ldots,i_{s,l})$, where each of the $l$ item-click events $i_{s,t}$ ($t=1\ldots l$) $\in$ $\mathcal{I}$, and $t$ denotes the position of the item $i_{s,t}$ in the session $s$.
The goal of SR is to predict the next item $i_{s,l+1}$ as the target class in an $m$-way classification problem by estimating the $m$-dimensional item-probability vector $\mathbf{\hat{y}}_{s,l+1}$ corresponding to the relevance scores for the $m$ items. The top-$K$ items are recommended based on the highest scores. 

To achieve this, we generally train SR models ($m$-class classifier model), denoted by $\mathcal{M}_r$ using cross-entropy (softmax) loss during training for estimating $\theta$ (parameters of $\mathcal{M}_r$) by minimizing 
    $\mathcal{L}(\hat{\mathbf{y}}) = - \sum_{j=1}^{m}\mathbf{y}_j log (\hat{\mathbf{y}}_j)$
over all training samples, where $\mathbf{y} \in \{0,1\}^m$ is a 1-hot vector with  $\mathbf{y}_k=1$ corresponding to the target class $k$.
In the literature, various models are employed to implement $\mathcal{M}_r$, e.g., RNNs \cite{hidasi2015session}, GNNs \cite{wu2018session}, Transformers \cite{kang2018self,xie2022contrastive}, etc. However, the challenge to enhance performance due to the sparsity of observed data causing popularity bias is overlooked.

\subsection{Guided Diffusion Model}
Following \cite{yang2024generate}, we focus on generating the counterfactual slate using the guided diffusion model. Typically, diffusion models generate via a two step process: i) \textit{Forward Process:} corrupting the input signal by adding noise
ii) \textit{Reverse Process:} learn to recover the original input signal from the noise. 
Given, the set of disjunct items, $\mathcal{I}$, each item $i_{s,j} \in \mathcal{I}$ is mapped to a  $d$ -dimensional representation or embedding vector $\bm{e}$. Thus, the session embedding is denoted as $\bm{e}_{s} = (\bm{e}_{s,1},\bm{e}_{s,2},....\bm{e}_{s,l})$.
Similar to \cite{yang2024generate}, we formulate the forward process  as a Markov chain of Gaussian transitions:
\begin{equation}
q(\bm{e}^t_{l+1}|\bm{e}^{t-1}_{l+1}) = \mathcal{N}(\bm{e}^t_{l+1};\sqrt{1-\beta_t}\bm{e}^{t-1}_{l+1},\beta_t\mathbf{I})
\end{equation}
with a variance scheduler $[\beta_1, \beta_2, ...\beta_T]$. 
\\
Further to guide the reverse (denoising) process, based on historical sessions, $\bm{e}_s$ is encoded with a Transformer Encoder: $\bm{c}_s = \textbf{T-enc}(\bm{e}_s)$, which conditions the reverse process as follows:
\begin{equation}
p_\theta(\bm{e}^{t-1}_{l+1}|\bm{e}^{t}_{l+1},\bm{c}_{s},t) = \mathcal{N}(\bm{e}^{t-1}_{l+1}; \bm{\mu}_\theta(\bm{e}^{t}_{l+1},\bm{c}_{s},t),\Sigma_\theta(\bm{e}^{t}_{l+1},\bm{c}_{s},t)),
\end{equation}
where, the architecture  of $\bm{\mu}_\theta(\bm{e}_{l+1}^{t}, \bm{c}_{s},t)$ is an MLP.

To manipulate the influence of the guidance signal $\bm{c}_{s}$,  $f_\theta(\bm{e}_{l+1}^t, \bm{c}_{s}, t)$ is modified to the following format:
\begin{equation}\label{eq:cfg}
    \tilde{f}_\theta(\bm{e}_{l+1}^t, \bm{c}_{s}, t) = (1 + w) \ f_\theta(\bm{e}_{l+1}^t, \bm{c}_{s}, t) - w \ f_\theta(\bm{e}_{l+1}^t, \Phi, t)
\end{equation}
where $w$ is a hyperparameter controlling the strength of $\bm{c}_{s}$:
\begin{equation}
     \bm{e}_{l+1}^{t-1} = \frac{\sqrt{\bar{\alpha}_{t-1} }\beta_t}{1 - \bar{\alpha}_t}  \tilde{f}_\theta(\bm{e}_{l+1}^{t}, \bm{c}_{s}, t)  + \frac{\sqrt{\alpha_t}(1-\bar{\alpha}_{t-1})}{1 - \bar{\alpha}_t}\bm{e}_{l+1}^t  + \sqrt{\tilde{\beta}_t} \mathbf{z},
\end{equation}
where, $\mathbf{z} \sim \mathcal{N}(\bm{0}, \bm{I})$. Finally, we obtain $\bm{e}^0_{l+1}$ by denoising a gaussian sample $\bm{e}_{l+1}^T \sim \mathcal{N}(\mathbf{0},\mathbf{I})$ for T steps. 
We retrieve the counterfactual slate of items $\hat{r}_j$ by obtaining the K-nearest items to the generated item embedding $\bm{e}_{l+1}^0$.
\subsection{Temporal SCM Response Model}
First we learn $\mathcal{M}_s$ as a temporal SCM model defined in \cite{gupta2024scm4sr}. Then, for a user $u$, we get $\bm{R}_t=\hat{r}_t$ via $\mathcal{M}_d$ i.e. counterfactual slate. 
The pretrained $\mathcal{M}_s$ is leveraged to
generate a counterfactual response $\bm{\hat{Y}}_t$. For this, we sample $\bm{\hat{\beta}}$ from $q_{\bm{\phi}}(\bm{\beta})$ and then compute the probability of an item at the $n^{th}$ position in $\hat{r}_t$ denoted by $\hat{r}_{t,n}$ by:
\vspace{-2mm}
\begin{equation}
\label{eq:response}
p_{\bm{Y}_t}(\hat{r}_{t,n}|\bm{UI}_{t}, \bm{R}_t = \hat{r}_t, \bm{\hat{\beta}}) = 
\frac{\exp{(\bm{UI}_{t,\hat{r}_{t,n}}+\mathbf{w}_{\hat{r}_{t,n}}\hat{\bm{\beta}}_{\hat{r}_{t,n}})}}{\sum_{k=1}^{K+1}\exp{(\bm{UI}_{t,\hat{r}_{t,k}}+\mathbf{w}_{\hat{r}_{t,k}}{\hat{\bm{\beta}}}_{\hat{r}_{t,k}})}}.
\end{equation} 
Finally, $\hat{\mathbf{Y}}_t$ is predicted by considering the top-$M$ items with the highest probability, where $M$ is number of positive clicked items in a given slate.

\begin{algorithm}[t]
	\caption{DCASR Framework \label{algo:scm}}
	\begin{algorithmic}
            \State Given observed data  $\mathcal{O} = \{\{u_j, r_{j,t},y'_{j,t}\}^l_{t=1}\}^{N_u}_{j=1}$, where, 
            \State $u_j$: individual user, $N_u$: Number of users
            \State $r_{j,t}$: corresponding recommended slate of items at time 't' \State $y'_{j,t}$: the user's response at timestep $t$, 
		\State   $\mathcal{M}_s$ and $\mathcal{M}_d$ are pre-trained based on the observed dataset $\mathcal{O}$.
        \State Initialize the counterfactual dataset $\mathcal{O}_c$ =  $\phi$
        \State Indicate number of running times = $N$ 
		\For{$i$ in $[0, N]$}  
            \State Select training sample
            \{$u, \{r_1,y_1\}, \{r_2,y_2\}, \dots, \{r_L,y_l\}$\}
            \State Select index $d'$ of the first clicked item slate to create a session 
            \State $s_c = \{i_{d'}\}$, $i_{d'}$ is clicked item from $r_{d'}$
            \For{$j$ in $[d'+1,l]$}            
                \State Obtain $\hat{r}_j$ from $\mathcal{M}_d$ 
                \State Obtain $\hat{y}_j$ from $\mathcal{M}_s$ using eq. \ref{eq:response}
                \If{$1 \in \hat{y}_j$} 
                   \State $s_c = s_c + i_j$
                \EndIf
            \EndFor    
            \State $\mathcal{O}_c$ = $\mathcal{O}_c \cup$ $s_c$
        \EndFor
        \State
        Optimize $\mathcal{M}_r$ using $\mathcal{O} \cup \mathcal{O}_c$
        \State Use $\mathcal{M}_r$ for obtaining final recommendations
	\end{algorithmic} 

\end{algorithm}
\section{Experimental Evaluation}
In this section, we conduct experiments to address the following research questions:
\begin{itemize}
\item \textbf{RQ1:} Does our proposed counterfactual data augmentation approach have the capability to improve the performance of the baseline SR models?
\item \textbf{RQ2:} Is the proposed approach better than the baseline Diffusion models for SR?
\item \textbf{RQ3:} How effective is  our proposed model at  handling popularity bias?
\end{itemize}
\subsection{Dataset and Evaluation Metrics}
We conduct our experiments on a simulated environment RecSim for online evaluation and also on a  real-world Diginetica dataset for offline evaluation.  \\
\textbf{RecSim:}
We conduct an online evaluation in a simulated envi-
ronment using an open-source user-behavior simulation
model RecSim \cite{ie2019recsim}, which has been recently considered for evaluating RS approaches \cite{garg2020batch,gupta2021causer,gupta2024scm4sr} . We modify RecSim to mimic a long-tailed distribution over item clicks (i.e., introducing popularity
bias) as follows: we consider i. two user types UT1 and
UT2, and 1k items such that UT1 and UT2 have high
propensity towards different subsets of items, and ii. the distribution across user types in the simulation model is 0.8 for UT1 and 0.2 for UT2, respectively. This results in items of interest for UT2  being less frequent in the historical session logs generated via a random agent
interacting with the simulator by simulating over 4k sessions of length five each. For evaluation we use a uniform distribution with UT1 and UT2 distributed as {0.5, 0.5} over 1k sessions (0.5k for each user type). We consider an
online evaluation metric CTR (Click Through Rate) i.e., percentage
of clicks across the test sessions and a popularity metric Average
Recommendation Popularity (ARP) as in [ 3 , 4 ] for online evaluation.
\\
\textbf{Diginetica:}
We use a large-scale real-world recommendation dataset
from CIKM Cup 2016 \cite{wu2018session} to evaluate the effectiveness of DCASR for offline re-ranking. We selected the top 10k popular clicked items and filtered out the sessions with session length $\le2$ greater than 10. The slate size was fixed 5. We selected 73,387/ 7,338/ 5,707 sessions for training/ validation/ testing based on chronological splits. The average length of sessions in training, validation and test was 3.11/ 2.56/ 3.08, respectively. \\
We use the standard offline evaluation metrics Recall$@K$ and Mean Reciprocal Rank (MRR$@K$) with $K=5$ as in \cite{wu2018session}, along with the popularity bias-related metric ARP. 
\begin{table*}[t]
\caption{Online and Offline evaluation on RecSim and Diginetica respectively. We report overall performance, as well as performance on user types, head, mid, and long-tail target items where the head, mid, and long-tail are categorized based on the popularity of the items such that each category has an equal number of sessions. Bold numbers are for the best performing methods.}
\label{tab:online}
\center
 \resizebox{\textwidth}{!}{
\begin{tabular}{@{}l|cccccc|cccccccccccccc@{}}
\toprule
\multicolumn{1}{c|}{\textbf{}} &
\multicolumn{6}{c|}{\textbf{RecSim (Online Evaluation)}} &
\multicolumn{12}{c}{\textbf{Diginetica (Offline Evaluation)}} \\ 
\cmidrule(lr){2-7}  
\cmidrule(lr){8-19} 
\textbf{Methods} &
\multicolumn{2}{c}{\textbf{UT1}} &
\multicolumn{2}{c}{\textbf{UT2}} &
\multicolumn{2}{c|}{\textbf{Overall}}&
\multicolumn{3}{c}{\textbf{Long-tail}} & \multicolumn{3}{c}{\textbf{Mid}} & \multicolumn{3}{c}{\textbf{Head}} &
\multicolumn{3}{c}{\textbf{Overall}} 
\\ \cmidrule(lr){2-3} \cmidrule(lr){4-5} \cmidrule(lr){6-7} 
\cmidrule(lr){8-10} \cmidrule(lr){11-13} \cmidrule(lr){14-16} \cmidrule(lr){17-19}
\multicolumn{1}{c|}{\textbf{}} &
\multicolumn{1}{c}{\textbf{CTR $\uparrow$}} & \multicolumn{1}{c}{\textbf{ARP $\downarrow$}} &
\multicolumn{1}{c}{\textbf{CTR $\uparrow$}} & \multicolumn{1}{c}{\textbf{ARP $\downarrow$}} &
\multicolumn{1}{c}{\textbf{CTR $\uparrow$}} & \multicolumn{1}{c|}{\textbf{ARP $\downarrow$}} &
\multicolumn{1}{c}{\textbf{R $\uparrow$}} & \multicolumn{1}{c}{\textbf{MRR $\uparrow$}} & \multicolumn{1}{c}{\textbf{ARP $\downarrow$}} & \multicolumn{1}{c}{\textbf{R $\uparrow$}} & \multicolumn{1}{c}{\textbf{MRR $\uparrow$}} & \multicolumn{1}{c}{\textbf{ARP $\downarrow$}} & 
\textbf{R $\uparrow$} & \textbf{MRR $\uparrow$} & \textbf{ARP $\downarrow$} & \textbf{R $\uparrow$} & \textbf{MRR $\uparrow$} & \textbf{ARP $\downarrow$} 
\\\midrule
DreamRec \cite{yang2024generate} & 73.7 &   0.38 & 69.22 & 0.39 & 71.50 & 0.38  & 5.91  & 5.51 & \textbf{0.03} & 4.56 & 4.01 & 0.03 & 4.11 & 3.07 & \textbf{0.03} & 4.68 & 3.97 & \textbf{0.03} \\
\midrule
\midrule
SRGNN \cite{wu2018session} & 83.10 & 0.73 & 41.42 & 0.73 & 62.26 & 0.73& 5.92 & 3.74 & 0.17 & 11.31 & 7.04 & 0.18 & 20.25 & 11.94 & 0.23 & 12.69 & 7.60 & 0.19 \\
+H-CASR \cite{wang2021counterfactual} & 83.68 & 0.71 & 30.47 & 0.70 & 57.07 & 0.70&  6.30 & 3.96 & 0.15 & 11.74 & 7.04 & 0.17 & 19.94 & 11.63 & 0.22 & 12.66 & 7.61 & 0.18 \\
+D-CASR \cite{wang2021counterfactual} & \textbf{83.75} & 0.70 & 31.96 & 0.68 & 57.85 & 0.70&5.98 & 3.67 & 0.17 & 11.47 & 7.25 & 0.18 & 20.10 & 11.95 & 0.23 & 12.52 & 7.57 & 0.19 \\ 
+M-CASR \cite{wang2021counterfactual} & 83.69 & 0.67 & 43.31 & 0.69 & 63.50 & 0.68 & 5.82 & 3.65 & 0.17 & 11.26 & 7.09 & 0.18 & \textbf{20.52} & 12.06 & 0.23 & 12.53 & 7.58 & 0.19 \\ 
 +SCM4SR-I \cite{gupta2024scm4sr} & 82.41 & 0.70 & 51.85 & 0.68 & 67.13 & 0.69 &  6.35 & 3.92 & 0.14 & \textbf{11.95} & 7.42 & 0.16  & \textbf{20.52} & \textbf{12.18} & \textbf{0.20} & 12.94 & 7.84 & 0.17 \\
 +SCM4SR-II \cite{gupta2024scm4sr} & 81.53 & 0.65 & 57.85 & 0.59 & 69.69 & 0.61 & 6.72 & 4.04 & 0.14 & 11.74 & 7.15 & 0.16 & 20.20 & 11.91 & 0.21 & 12.90 & 7.70 & 0.17 \\
 \textbf{+DCASR-I} & 76.39
 & 0.41
  & 66.43
  & 0.42 & 71.41 & 0.41 & \textbf{7.10} & \textbf{5.21}  &  0.14 & \textbf{11.95} & \textbf{7.69} & 0.16 & 20.10 & 11.87 & 0.21 & \textbf{12.97} & \textbf{8.2} & 0.17 \\
 \textbf{+DCASR-II} & 75.63 & 0.44  & \textbf{70.63} & 0.44 & \textbf{73.13} & 0.43 & 6.67 & 4.66 & 0.14 & 11.63 & 7.48 & 0.16 & 20.31 & 11.91 & 0.21 & 12.78 & 7.95 & 0.17 \\
 
\midrule
\midrule
NISER \cite{gupta2019niser}  & 81.18 & 0.59 & 57.89 & 0.60 & 69.54 & 0.59& 6.30 & 4.42 & 0.14 & 10.08 & 7.00 & 0.17 & 19.83 & 12.01 & 0.22 & 12.07 & 7.81 & 0.18\\ 
+H-CASR \cite{wang2021counterfactual} & 81.67 & 0.63 & 53.68 & 0.64 & 67.67 & 0.59& 6.40 & 4.50 & 0.15 & 10.67 & 7.14 & 0.17 & 19.67 & 11.99 & 0.21 & 12.25 & 7.88 & 0.18\\
+D-CASR \cite{wang2021counterfactual} & \textbf{81.87} & 0.62 & 52.92 & 0.63 & 67.39 & 0.63  & 6.35 & 4.49 & 0.15 & 10.14 & 7.02 & 0.17 & 19.88 & 11.99 & 0.22 & 12.12 & 7.83 & 0.18 \\
+M-CASR \cite{wang2021counterfactual} & 81.28 & 0.60 & 56.87 & 0.62 & 69.07 & 0.61& 5.02 & 3.46 & 0.17 & 9.45 & 6.33 & 0.19 & 19.30 & 11.91 & 0.23 & 11.25 & 7.23 & 0.20 \\
+SCM4SR-I \cite{gupta2024scm4sr} & 80.07 & 0.58 & 58.65 & 0.58 & 69.36 & 0.58 &  6.72 & 4.57 & 0.13 & 11.85 & 7.32 & 0.15 & 20.84 & 11.54 & 0.20 & 13.14 & 7.81 & 0.14  \\
+SCM4SR-II \cite{gupta2024scm4sr} & 79.68 & 0.59 & 60.79 & 0.59 & 70.23 & 0.59 &  7.26 & 4.86 & 0.14 & 10.78 & 6.30 & 0.16 & \textbf{21.96} & \textbf{12.43} & 0.21 & 13.34
 & 7.86 & 0.15\\
 \textbf{+DCASR-I} & 78.45 & 0.44 & \textbf{66.51}  & 0.41 & \textbf{72.48} & 0.42 & 7.68 & 5.57 & 0.13 & 12.22 & 7.49& 0.15 & 20.52 & 12.21  & 0.20  & 13.38
 & 8.38 & 0.16\\
 \textbf{+DCASR-II} & 78.15 & 0.43 & 65.47 & 0.38 & 71.81  & 0.40  & \textbf{7.90} &  \textbf{5.73} & 0.13 & \textbf{12.54} & \textbf{8.18} & 0.15 & 20.31 & 11.98 & 0.20  & \textbf{13.47}
 & \textbf{8.58} & 0.16 \\
\midrule
\midrule
CauSeR \cite{gupta2021causer} & \textbf{76.65} & 0.40 & 69.00 & 0.41 & 72.83 & 0.40 &  6.40 & 4.57 & 0.13 & 10.99 & 7.34 & 0.15 & 20.31 & 11.51 & 0.20 & 12.57 & 7.81 & 0.16 \\ 
+H-CASR\cite{wang2021counterfactual} & 75.24 & 0.40 & 71.05 & 0.40 & 73.15 & 0.40 & 6.14 & 4.36 & 0.16 & 11.15 & 7.22 & 0.18 & 21.91 & 12.23 & 0.23 & 13.07 & 7.94 & 0.19 \\
+D-CASR \cite{wang2021counterfactual} & 75.79 & 0.40 & 71.60 & 0.41 & 73.70 & 0.41 & 6.40 & 4.54 & \textbf{0.13} & 10.94 & 7.25 & \textbf{0.15} & 20.10 & 11.36 & \textbf{0.20} & 12.48 & 7.72 & \textbf{0.16}\\
+M-CASR \cite{wang2021counterfactual} & 73.43 & 0.40 & 74.40 & 0.40 & 73.91 & 0.40& 4.27 & 2.90 & 0.19 & 9.93 & 6.38 & 0.20 & 21.11 & 11.79 & 0.24 & 11.77 & 7.02 & 0.21 \\
+SCM4SR-I \cite{gupta2024scm4sr} & 70.43 & 0.35 & 76.79 & 0.35 & 73.61 & 0.35 & 6.40 & 4.05 & 0.15 & 12.33 & 7.48 & 0.17 & 21.16 & 12.09 & 0.20 & 13.51 & 7.92 & 0.18 \\
+SCM4SR-II \cite{gupta2024scm4sr} & 72.20 & 0.39 & 75.88 & 0.36 & \textbf{74.04} & 0.37  & 6.30 & 4.42 & 0.15 & 11.74 & 7.20 & 0.17 & 21.75 & \textbf{13.02} & 0.22 & 13.46 & 8.28 & 0.18\\
\textbf{+DCASR-I} & 68.89
 & 0.32  & 76.76 & 0.32  & 72.83  & 0.32  & 7.15 & 5.43 & 0.14 & 12.81 & 8.13 & 0.16 & 21.75 & 12.85 & 0.22 & 13.84 & 8.58 & 0.16 \\
\textbf{+DCASR-II} &  68.52 & \textbf{0.30} & \textbf{77.63} & \textbf{0.30} & 73.07  & \textbf{0.30} & \textbf{7.58} & \textbf{5.45} & 0.13 & \textbf{13.23} & \textbf{8.21} & 0.15 & \textbf{21.85}  & 12.29 & 0.20  & \textbf{14.07} & \textbf{8.59} & 0.18 \\
\bottomrule
\end{tabular}
}
\end{table*}

\subsection{Baselines}
To show the efficacy of the \textbf{DCASR} framework, we compare it with several state-of-the-art methods from the literature. 
First, we consider three distinct SR models to model $\mathcal{M}_r$: i) \textbf{SRGNN \cite{wu2018session}} is a vanilla SR model that models each session as a graph, ii) \textbf{NISER \cite{gupta2019niser}} further introduces an item and session-graph representation learning mechanism to handle popularity bias at the data level, iii)
  \textbf{CauSeR \cite{gupta2021causer}} is a causal inference based  SR model which handles popularity bias at the data-generation and deep-learning model level.

Further, we compared \textbf{DCASR} with i. vanilla SR model (as mentioned above), ii. three variants from \cite{yang2021top}, i.e., Heuristic (+H-CASR), Data-oriented (+D-CASR), and Model-oriented (+M-CASR), 
iii. two variants from \cite{gupta2024scm4sr}, i.e., +SCM4SR-I and +SCM4SR-II.
In addition to the aforementioned SR models, we compare our framework with the recently proposed guided diffusion model for Sequential recommendations i.e. \textbf{DreamRec} \cite{yang2024generate}.
\subsection{Implementation Details}
We use Recall$@1$ as the performance metric for hyperparameter selection on offline validation data for all the approaches.  For training $\mathcal{M}_d$ and $\mathcal{M}_s$, we search over the same parameters as in \cite{yang2024generate} and \cite{gupta2024scm4sr} respectively.
  For NISER, we use the same parameters \cite{gupta2019niser} i.e., scaling factor $= 16.0$. For CauSeR, we grid-search over $\alpha$ in \{$0, 1.0, 2.5, 5.0, 7.5, 10.0$\}, $\beta$ in $\{0, 0.001, 0.01, 0.05, 0.1, 0.5, 1.0\}$. The best parameters on the validation set $\alpha=5.0$, $\beta=0.5$ and $\alpha=5.0$, $\beta=0.01$ for RecSim and DN, respectively. 
 Slate size is $K=3$ and $K=5$ for RecSim and DN respectively. 
\subsection{Results and Observations}
From table \ref{tab:online}, 
DACSR shows significant performance improvement over the competitive state-of-the-art methods. 
\subsubsection{RQ1: Overall Comparison} 
DCASR variants perform significantly better than the respective SR baselines SRGNN, NISER and CauSER on both the datasets. For instance, DCASR achieved the CTR of $73.43$ vs $62.26$ for SRGNN ,$72.48$ vs $69.54$ for NISER and $72.83$ vs $73.07$ for CauSeR. We observe DCASR performs consistently better in terms of both Recall$@5$ and MRR$@5$ on Diginetica dataset. Moreover, compared to the current most advanced counterfactual augmentation framework, SCM4SR, our method have achieved considerable gains on both the datasets.
\subsubsection{RQ2: Comparison with diffusion generative model} We observe that the DreamRec model when trained as a SR model under performs DCASR on both RecSim and Diginetica datasets in terms of CTR, Recall$@5$ and MRR$@5$ respectively. However, from table \ref{tab:online}, it can be observed that the DreamRec outperforms all the methods in terms of ARP for Diginetica dataset. This demonstrates that the DreamRec model effectively generates less popular items. The DCASR framework utilizes this capability of the DreamRec model for generating counterfactual/alternate slates of items consisting of less popular items. 
\subsubsection{RQ3: Handling Popularity bias}
As mentioned above, the DCASR framework is certainly capable of handling popularity bias by generating more meaningful counterfactual sessions as compared to the counterfactual sessions generated by SCM4SR.
From table \ref{tab:online}, it can be observed that the DCASR method alleviates the popularity bias problem, by achieving up to $20\%$ gain in Recall$@5$ and $13\%$ gain in CTR on long-tail in Diginetica and UT2 in the RecSim dataset, respectively. 

\section{Conclusion}
In this work, we proposed a guided diffusion-based counterfactual data augmentation framework for SR (DCASR) that outperforms competing methods while generating relevant counterfactual sessions by combining the capabilities of generative diffusion models and causal models for SR.
We conducted experiments on a simulated environment RecSim and real-world Diginetica dataset and evaluated our approach against three distinct SR models and state-of-the-art augmentation frameworks. Experiment results indicate that DCASR outperforms existing baseline methods especially for less popular target items by generating relevant counterfactual sessions.

\bibliographystyle{ACM-Reference-Format}
\balance
\bibliography{recsys24}


\end{document}